\documentclass[reprint, aps, pre,  superscriptaddress]{revtex4-2}

\usepackage{amsmath}
\usepackage{graphicx}
\usepackage{epstopdf}
\usepackage{color}
\usepackage{amssymb}
\usepackage{amsmath}
\usepackage{tabularx}
\usepackage{bm}
\usepackage{hyperref}
\usepackage{ltxtable}
\usepackage{natbib}
\usepackage{longtable}
\usepackage[normalem]{ulem}
\usepackage{float}
\usepackage{graphicx}
\usepackage{feynmf}
\usepackage{tikz}
\usepackage[compat=1.1.0]{tikz-feynman}

\newcommand{\la}{\left\langle}
\newcommand{\ra}{\right\rangle}
\newcommand{\be}{\begin{equation}}
\newcommand{\ee}{\end{equation}}
\newcommand{\bse}{\begin{subequations}}
\newcommand{\ese}{\end{subequations}}
\newcommand{\bea}{\begin{eqnarray}}
\newcommand{\eea}{\end{eqnarray}}
\newcommand{\ba}{\begin{array}}
\newcommand{\ea}{\end{array}}

\begin{document}

\title{Contrasting   thermodynamic and hydrodynamic entropy}
\author{Mahendra K. Verma}
\email{mkv@iitk.ac.in}
\affiliation{Department of Physics, Indian Institute of Technology Kanpur, Kanpur 208016, India}
\author{Rodion Stepanov}
\email{rodion@icmm.in}
\affiliation{Institute of Continuous Media Mechanics, Korolyov 1, 614013, Perm, Russia}
\affiliation{Perm National Research Polytechnic University, Komsomolskii av. 29, 614990 Perm, Russia}
 \author{Alexandre Delache}
 \email{alexandre.delache@ec-lyon.fr}
 \affiliation{Ecole Centrale de Lyon, CNRS, Universite Claude Bernard Lyon 1, INSA Lyon, LMFA, UMR5509, 69130, Ecully, France}
  \affiliation{Jean Monnet University, 42100 Saint-Étienne, France}
\date{\today}

\begin{abstract}
In this paper, using \textit{hydrodynamic entropy} we quantify the multiscale disorder in Euler and hydrodynamic turbulence. These examples illustrate that the hydrodynamic entropy is not extensive because it is not proportional to the system size. Consequently, we cannot add hydrodynamic and thermodynamic entropies, which measure disorder at macroscopic and microscopic scales, respectively. In this paper, we also discuss the hydrodynamic entropy for the time-dependent Ginzburg-Landau equation and Ising spins. 
\end{abstract}
\maketitle

\section{Introduction}

The universe is a fascinating mix of order and disorder. For example, the paramagnet transforms to a ferromagnet below the critical temperature. Thermodynamic entropy (TE in brief) quantifies disorders in these systems. \citet{Boltzmann:book_chapter} and \citet{Gibbs:book:StatMech} showed that TE has a microscopic origin, and it is related to the configurations of spins and atoms of the system~\cite{Huang:book:SM}. The microscopic entropy is often referred to as   \textit{\em Boltzmann entropy}, \textit{BE} in short.   The phase transitions in such systems have been explained using equilibrium statistical physics~\cite{Landau:book:StatMech,Wilson:PR1974,Huang:book:SM}. For example, a ferromagnet has a long-range order (LRO), with a nonzero average of the order parameter~\cite{Huang:book:SM}. Refer to Appendix A for a brief discussion on TE, BE, and Gibbs entropy.

Nonequilibrium systems have another kind of disorder.   { For example, the correlations of the order parameter (here, velocity field)  in thermal-convection rolls~\cite{Chandrasekhar:book:Instability}, Jupiter's red spot, and Earth's Hadley shells ~\cite{Vallis:book} are dynamic, and they differ from those in a equilibrium systems.  A typical route to turbulence in nonequilibrium systems follows the following path: instability, pattern formation, spatio-temporal chaos, and then turbulence~\cite{Manneville:book:Instabilities}.  In the following discussions we show that BE is  inadequate for describing disorder in nonequilibrium systems. Instead, \textit{hydrodynamic entropy} (HE in short)~\cite{Verma:PRF2022} is a  quantifier for the disorder in a variety of nonequilibrium systems.

In the following discussion we  focus on the  disorder in nonequilibrium multiscale systems, namely in \textit{Euler turbulence} and in \textit{thermal convection}.  Variations in TE is proportional to the heat production, which is absent in an Euler flow   due to lack of viscosity. Therefore, TE of an Euler flow is constant~\cite{Landau:book:Fluid}. However, Euler flow exhibits a nontrivial temporal evolution and finite energy flux when the  flow starts with large-scale structures~\cite{Cichowlas:PRL2005}.    In a recent work, \citet{Verma:PRF2022}  showed that HE  describes the  disorder in  Euler turbulence quite well. For an ordered initial condition, 3D Euler equation evolves from order to disorder or HE increases with time.  However, for several coherent initial conditions, 2D Euler turbulence  becomes more ordered with time, or HE decreases for a finite time duration. Thus, HE provides valuable insights and quantitative descriptions for Euler turbulence.

\begin{figure}[h]
	\centering
	\includegraphics[width=1\columnwidth]{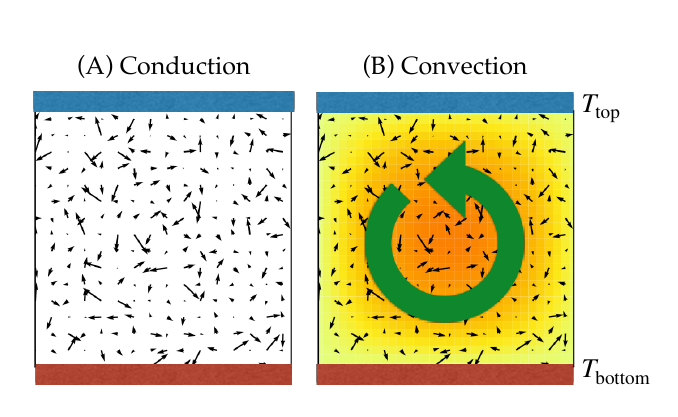} 
	\caption{ A box of ideal gas is being heated from below and cooled from top. Here we illustrate the velocities of the molecules for  (A) Conduction state  and	(B) Convection state. In (A), the molecules move with random thermal speed, whereas in (B) the molecules  have an additional  velocity due to the convection roll. The convection state is more ordered than the conduction state; this contrast can be quantified using hydrodynamic entropy.} 
	\label{fig:convection}
\end{figure}


Now we turn to thermal convection, for which   Prigogine provided valuable insights in terms of dissipative structures~\cite{Prigogine:Science1978}.     Here, a fluid is confined between two plates, with the bottom plate at $T_\mathrm{bottom}$ and the top plate at $T_\mathrm{top}$, and $T_\mathrm{bottom} > T_\mathrm{top}$.  See Fig.~\ref{fig:convection} for an illustration. For a temperature difference ($T_\mathrm{bottom} - T_\mathrm{top}$) below a critical value $(\Delta T)_c$,   heat is transported by conduction in the stationary fluid. However,  convective rolls are born when the temperature difference exceeds $(\Delta T)_c$.  A typical convection experiment  with air has the temperature difference  varying from 10 to 100 Celsius. In such a system, the hydrodynamic velocity  of the convection rolls is $O(1~\mathrm{cm/s})$, which is much smaller  than the thermal speed ($\sim$ 300 m/s). As shown in Fig.~\ref{fig:convection}, the molecules in the conduction state move  randomly with the thermal velocity, whereas the molecules in the convection state have an additional hydrodynamic (coherent) velocity, which is much smaller than the thermal speed. 

Near the onset of convection, the average temperatures of the conduction and convection states are nearly the same,  leading to approximately similar Boltzmann entropies (BEs) for the two configurations [see Eq.~(\ref{eq:S_thermo_gas})].   BE describes the disorder for the conduction state quite well, but it fails to capture the order of the convective roll [see Fig.~\ref{fig:convection}(B)].  Similar features exist in Earth's atmosphere too. Here, the hydrodynamic velocity, which is $O(1~\mathrm{m/s})$, is much smaller than the thermal speed.  These observations mandates construction of  a new formula to quantify the hydrodynamic order.  Prigogine~\cite{Prigogine:Nobel}  highlights this aspect as follows, ``the  canonical  distribution  would  assign  almost  zero  probability  to  the  occurrence  of  B\'{e}nard  convection.  Whenever  new  coherent  states  occur  far  from equilibrium,  the  very  concept  of  probability,  as  implied  in  the  counting  of number   of   complexions, breaks down".    As we show in this paper, the framework of  \textit{hydrodynamic entropy} overcomes this deficiency of equilibrium statistical mechanics.

Quantification of disorder and correlations in  biological, economic, and social systems remains a challenge. However, there are interesting applications of Shannon's entropy~\cite{Shannon:BELL1948}  to cities, ecology, and income distributions~\cite{Golan:PNAS2022,Mishra:Risk2019,Rajarama:Complexity2017,Zhou:Entropy2013}. Other  leading applications of entropy measures include   electroencephalogram (EEG)~\cite{Deco:Cell_Reports2020}, electrocardiogram (ECG)~\cite{Asgharzadeh:Bio2020}, and chaos~\cite{Clark:PRF2020}.  Interestingly, most of the above systems---cities, finance, income distribution, EEG, ECG---are multiscale, for which HE could provide valuable measures and insights.  In fact, some works on EEG~\cite{Deco:Cell_Reports2020}, ECG~\cite{Asgharzadeh:Bio2020}, and chaos theory~\cite{Clark:PRF2020} employ Fourier transforms, as in HE, for entropy quantification.

In this paper, we define HE, contrast it with thermodynamic entropy, and then employ HE to variety of systems---Euler turbulence, hydrodynamic turbulence, Ising spins, and coarsening. The  \textit{hydrodynamic entropy maxima} of  Euler turbulence and Ising spins are proportional to the logarithms of the respective grid sizes, indicating that HE is nonextensive. We show that the HE of  hydrodynamic turbulence is between 3 and 4, and it is  independent of the grid size.  Our results appear quite general and illuminating, and we expect similar results to hold for other nonequilibrium systems.}


 The outline of the paper is as follows. We define hydrodynamic entropy in Section~\ref{sec:HE_defn}, and compute  hydrodynamic entropies for Euler and fluid turbulence in Sections~\ref{sec:HE_Euler} and \ref{sec:HE_turbulent} respectively. Section~\ref{sec:HE_statmech} discusses the hydrodynamic entropies for the Ising spins and time-dependent Ginzburg-Landau equation. We conclude in Section~\ref{sec:conclusions}. In Appendix~\ref{sec:Appendix}, we summarize the Boltzmann entropy, Gibbs entropy, Shannon entropy, von Neumann entropy, and Tsallis entropy in order to compare them with hydrodynamic entropy.

\section{Definition of Hydrodynamic entropy}
\label{sec:HE_defn}

In this section, we define hydrodynamic entropy (HE), which is primarily suitable for multiscale nonequilibrium systems~\cite{Aubry:JSP1991, Verma:PRF2022}.  The hydrodynamic entropy is computed using the energy spectrum as follows.  { For a scalar quantity  $\psi$, the modal energy is defined as
\be
E({\bf k}) = \frac{1}{2} |\psi({\bf k})|^2,
\ee
where $\psi({\bf k})$ is the Fourier transform of $\psi({\bf x})$.  The corresponding modal energy for vector field (e.g., velocity field) ${\bf u}$ is 
\be
E({\bf k}) = \frac{1}{2} |{\bf u}({\bf k})|^2.
\ee }
Nonequilibrium systems  driven at large scales create  structures with different energies, among which  the large-scale structures  dominate the order or correlations. Following this idea, \citet{Verma:PRF2022} proposed that the probability  $p_{\bf k}$ for wavenumber ${\bf k}$ is proportional to its modal energy $E({\bf k})$. That is,
\be
p_{\bf k} = \frac{E({\bf k})}{\sum_{\bf k} E({\bf k}) }
\label{eq:p_k}
\ee
with  $\sum_{\bf k} p_{\bf k}  = 1$, and
HE  is
\be
S_H = - \sum_{\bf k}  p_{\bf k} \log_2 p_{\bf k}.
\label{eq:S_H}
\ee
In this measure, larger the energy of a mode, larger its contribution to the  HE. { For isotropic systems, as in isotropic hydrodynamic turbulence, $E({\bf k})$ depends on $k$.  For such systems, we  redefine $p_{\bf k} $ and $S_H$ of Eqs.~(\ref{eq:p_k}, \ref{eq:S_H}) as follows:
\bea
p_{ k} & = & \frac{E({ k})}{\sum_{ k} E({ k}) } ,
\label{eq:pk_shell} \\
S_H & = & - \sum_{ k}  p_{ k} \log_2 p_{ k},
\label{eq:SH_shell}
\eea 
where $E(k) =\sum_{k-1 < |{\bf k'}| \le k} E({\bf k'}) $ is the shell spectrum for unit-width shell of radius $k$. Equations~(\ref{eq:pk_shell}, \ref{eq:SH_shell}) lead to considerable simplification in the computation of $S_H$ for isotropic systems. }
Interestingly, HE can be computed for a given  snapshot, a convenient feature using which we can compute the entropy time series for a system.  Thus, HE can yield valuable insights into the evolution of dynamical systems.

{
Let us contrast  HE with  the thermodynamic entropy (TE) or Boltzmann entropy (BE).  Application of BE hinges on the \textit{a priori  probability postulate} and \textit{ergodic hypothesis}  of statistical mechanics \cite{Huang:book:SM}.    Classical and quantum statistical  mechanics typically deal with Hamiltonian systems whose energies are conserved. We consider a closed system with $N$ interacting particles that moves in  $6N$ dimensional phase space. A \textit{microstate} of the system contains information on the positions and velocities of all the particles, but  a \textit{macrostate} is a combination of many microstates with similar configurations.   Boltzmann related the thermodynamic entropy to all  the microstates for a given macrostate~\cite{Boltzmann:book_chapter,Lebowitz:PA1993}.  See Appendix A.

It is conjectured that the systems of statistical mechanics reach thermal equilibrium, or \textit{thermalize}, asymptotically.  According to  \textit{ergodic hypothesis},  a system under equilibrium covers  the  available phase space (satisfying energy conservation) such that the temporal average of a quantity equals its ensemble average.  
Intuitively, the  \textit{a-priori  probability postulate} originates from this observation.  In addition, in the equilibrium state, BE is maximum, and the energy is equipartitioned among all the modes. Proving that a system is ergodic is quite complex, but it is much simpler to show thermalization for a system.  In this context,  \citet{Cichowlas:PRL2005} showed that three-dimensional (3D) Euler turbulence (energy conserving system) reaches equilibrium asymptotically. However, for some initial conditions, two-dimensional (2D) Euler turbulence does not reach equilibrium~\cite{Verma:PRF2022}. Using numerical simulations, \citet{Orszag:CP1973} analyzed a five-mode energy conserving model, resembling Euler equation, and showed that this model thermalizes and becomes ergodic. Thus, behaviour of the Euler  equation is both consistent and inconsistent depending on the initial condition.

Note, however, that  many natural systems are driven and dissipative. Hence, the validity of the thermalization conjecture remains uncertain for such systems.  The total energy is not conserved in such systems, but it fluctuates around the system's steady-state average~\cite{Pope:book}. In addition, the energy is typically concentrated at large scales, rather than being equipartitioned among the available modes.   Thus, nonequilibrium steady states   are very different from the equilibrium states of statistical mechanics. 

As we show later in this paper, HE is more appropriate than BE for quantifying order in nonequilibrium systems.  In Rayleigh-B\'{e}nard convection, the convection rolls  contribute to the HE. In driven-dissipative nonequilibrium system---turbulence, earthquakes, financial market, galaxies---the energy decreases when we go from large scales to small scales~\cite{Pope:book,Verma:EPJB2019}. The disorder in such systems can be satisfactorily quantified using  HE, as we show in subsequent sections of this paper.  
}

Note, however, that some unforced and dissipation-less hydrodynamic systems, e.g., Euler equation, reach equilibrium where the total energy is equipartitioned among all the available Fourier modes; the equilibrium state has the maximum HE.  We also show that  unlike BE, HE  is not extensive.

In the following sections we will compute HE for Euler and hydrodynamic turbulence,  the Ising spins, and  $\phi^4$ theory.  Using these examples, we will  contrast HE with  TE.  For example, we  illustrate that  HE is  nonextensive.

\section{Hydrodynamic Entropy for Euler Turbulence}
\label{sec:HE_Euler}

Euler equation that describes fluid flow with zero viscosity and no external force is~\cite{Lee:QAM1952,Kraichnan:JFM1973,Lesieur:book:Turbulence}
\be
\frac{\partial {\bf u}}{\partial t} + {\bf u \cdot \nabla u} = -\nabla p ,
\ee
where ${\bf u}, p$ are the velocity and pressure fields respectively. We assume the fluid to be incompressible ($\nabla \cdot  {\bf u} = 0$). 
For an ordered initial condition (e.g., Taylor-Green vortex),  3D Euler turbulence evolves from order to disorder, and it reaches an equilibrium state asymptotically~\cite{Cichowlas:PRL2005}.  But, the thermodynamic entropy (TE) of this system remain constant due to an absence of viscosity or heat production~\cite{Landau:book:Fluid}.    Note that Euler equation does not include any thermal or dissipative component, as in Navier-Stokes equation (see Sec.~\ref{sec:HE_turbulent}).  Due to these reasons, TE cannot capture the above variations in the disorder. As we illustrate below,  HE  successfully captures the evolution towards disorder in Euler turbulence~\cite{Verma:PRF2022}.  

\citet{Lee:QAM1952} and \citet{Kraichnan:JFM1973} derived the equilibrium solutions of Euler turbulence. In the absence of \textit{kinetic helicity } (${\bf u \cdot (\nabla \times u})$), the velocity correlation in the equilibrium state is delta-correlated in space and time that leads to statistical equipartition of energy among all the Fourier modes~\cite{Verma:PTRSA2020}. Euler turbulence is often simulated using a finite grid in Fourier space, hence it is also referred to as {\em truncated Euler turbulence}.  If $G$ is the number of grid points in each direction, then the total number of grid points in 3D is  $M = G^3$. Since the energy is equipartitioned, 
\be
\la E({\bf k}) \ra = \frac{1}{2} \la |{\bf u}({\bf k})|^2 \ra = \mathrm{const},
\ee
which leads to 
\be 
p_{\bf k}=  \frac{ \la E({\bf k}) \ra }{\sum_{\bf k} E({\bf k}) } =\frac{1}{M}.
\ee
 Therefore, using Eq.~(\ref{eq:S_H}) we derive  the HE  of the equilibrium state as
\be
S_H = \sum \frac{1}{M}  \log_2 M  =  \log_2 M  .
\label{eq:S_H_Euler}
\ee
For ordered initial conditions, 3D Euler flows evolve from order to disorder. The asymptotic state of 3D Euler turbulence has the maximum entropy, which is $\log_2 M$, but the intermediate states have HE less than $\log_2 M$.  The evolution for 2D Euler turbulence is more complex. Random delta-correlated initial conditions lead to equilibrium states with maxium HE, but some ordered initial conditions yield nonequilibrium asymptotic states, for which HE decreases for some duration (or the system becomes more ordered)~\cite{Verma:PRF2022}.

As shown in Eq.~(\ref{eq:S_H_Euler}), the maximum HE of Euler turbulence   is proportional to $\log_2 M$, not  to $M$,  as in the BE of Ising spin (see  Appendix A). Hence, HE is not proportional to the system size. Therefore, the HE of  Euler turbulence is nonextensive.     We remark that the dissipation-less Burgers  and Korteweg–De Vries (KdV) equations exhibit similar properties~\cite{Verma:PRE2022}. That is, they approach equilibrium states whose HE is proportional to $\log_2 M$, where $M$ is the grid size.

Thus, HE successfully captures the disorder in Euler turbulence. In contrast, the TE  remains constant during its evolution.

\section{Hydrodynamic Entropy of a Turbulent Flow}
\label{sec:HE_turbulent}

The incompressible Navier-Stokes equation, an extension of Euler equation with kinematic viscosity $\nu$ and external force ${\bf F}_\mathrm{ext}$, is
\be
\frac{\partial {\bf u}}{\partial t} + {\bf u \cdot \nabla u} = -\nabla p  + \nu \nabla^2 {\bf u}+{\bf F}_\mathrm{ext}. 
\ee
We assume the density of the fluid to be unity. The above equation exhibits turbulent behaviour for large Reynolds number $\mathrm{Re} = UL/\nu$, where $U,L$ are the large-scale velocity and length scales respectively. In this section, we will compute the hydrodynamic entropy of a turbulent flow.

In a turbulent system,  fluid structures contribute to the \textit{coherent hydrodynamic energy}, whereas   molecular motion at the microscopic scales contribute to the \textit{incoherent thermal energy}.  Here, the large-scale hydrodynamic energy cascades to small scales, where the hydrodynamic energy is  converted to the thermal energy of the molecules~\cite{Pope:book}.  Thus, the fluid gets hotter at the expense of hydrodynamic energy of the flow.  The  TE captures the disorder at the microscopic scales, whereas HE captures the macroscopic disorder of the hydrodynamic structures.  

In the following discussion, we \textit{estimate} the HE of a turbulent flow that follows Kolmogorov's  energy spectrum~\cite{Kolmogorov:DANS1941Structure,Pope:book}, i.e.,
\be 
E(k) = K_\mathrm{Ko} \epsilon^{2/3} k^{-5/3},
\ee
where $\epsilon$ is the energy flux, and $K_\mathrm{Ko}$ is Kolmogorov's constant. { The derivation of HE  is simplified when we divide the wavenumber sphere into logarithmically-binned $G$ shells. } Since Kolmogorov's theory of turbulence follows a powerlaw, we divide the wavenumber sphere into shells whose radii $k_n$'s are given by   
\be
k_n L = \lambda^n,
\label{eq:kn}
\ee
where $n$ (ranging from 1 to $G$) is the shell index, $L$ is the box size, and $\lambda$ is a parameter which is greater than 1. For simplicity, we ignore the dissipation range, and assume the minimum and maximum shell indices to be  1 and $G$ respectively. { The energy of the $n^\mathrm{th}$  shell is
\be
E_n = K_\mathrm{Ko} \int_{k_n}^{k_{n+1}} dk \epsilon^{2/3} k^{-5/3} \approx 
\frac{3 }{2} K_\mathrm{Ko} \epsilon^{2/3} k_n^{-2/3} [1-\lambda^{-2/3}].
\ee
We simplify our calculation by setting $(3/2) K_\mathrm{Ko} [1-\lambda^{-2/3}] \approx 1$ that leads to 
\be
E_n = \epsilon^{2/3} k_n^{-2/3}.
\ee

The above spectrum is similar to that in the \textit{shell model}~\cite{Gledzer:DANS1973,Biferale:ARFM2003}. Similarly, the total kinetic energy of the system is approximately
\be
E = \int_{k_0}^{G} dk E(k) \approx \int_{k_0}^{\infty} dk E(k) \approx  \epsilon^{2/3} L^{2/3}.
\ee
}
We estimate  the probability $p_n$ using
\be
p_n = \frac{E_n}{E} = A (k_n L)^{-2/3} = A \lambda^{-\alpha n},
\label{eq:pn}
\ee
where $\alpha = 2/3$, and $A$ is the prefactor, which is chosen so as to conserve probability. { We have lumped  the prefactors of earlier equations into   $A$. }
The constraint $\int dn p_n = 1$ yields
\be
A = 1/\int_1^G dn \lambda^{-\alpha n}
= \frac{\alpha \ln \lambda}{\lambda^{-\alpha} - \lambda^{-\alpha G}}.
\ee
{The choice of  $k_n$  of Eq.~(\ref{eq:kn}) plays an important role in making the constant $A$ dimensionless. For a linearly spaced $k_n$, the probability $p_n = D (k_n L)^{-2/3}/k_n$, where $D$ is a constant with dimension  $L^{-1}$.  Thus, logarithmically-binned shells simplify our $S_H$ computation significantly. }

\begin{figure}[h]
	\centering
	\includegraphics[width=1\columnwidth]{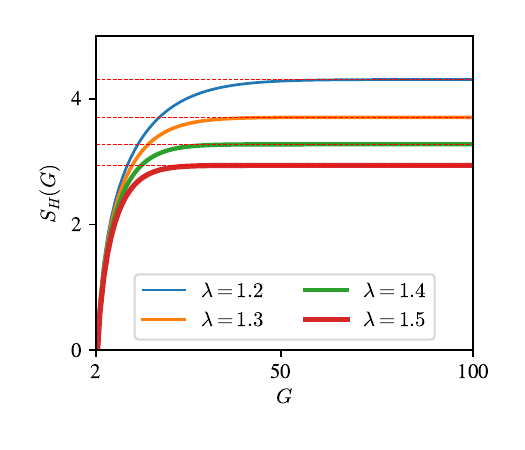} 
	\caption{Plots of hydrodynamic entropies computed using   the analytical formula [Eq.~(\ref{eq:S_H_anlytical})] for $\alpha = 2/3$ and $\lambda=1.2, 1.3, 1.4, 1.5$ using colored lines of increasing thicknesses. The asymptotic values [Eq.~(\ref{eq:S_turb})] for the four  $\lambda$'s are 4.31, 3.71, 3.26, and 2.94 respectively (shown using horizontal dashed lines). 	} 
	\label{fig:hydro_model}
\end{figure}

{ Using $p_n$ of Eq.~(\ref{eq:pn}), we compute  the hydrodynamic entropy as
\bea
S_H & = & -\sum_n p_n \log_2 p_n \nonumber \\
& \approx & -\int_1^G dn p_n \log_2 p_n 
 \nonumber \\
 & = & -\int_1^G dn A \lambda^{-\alpha n} \left[\log_2 A - \alpha n \log_2 \lambda \right]
 \nonumber \\ 
 & = & \left[ A \frac{\log_2\lambda}{\ln \lambda} -A \log_2 A \right] \left[ \frac{\lambda^{-\alpha}-\lambda^{-\alpha G}}{\alpha \ln \lambda} \right]   \nonumber \\ 
 &&+  \frac{\log_2\lambda}{\ln \lambda} A \lambda^{-\alpha G}(1-G).
\label{eq:S_H_anlytical}
\eea
In the limit $G \to \infty$, $A \rightarrow  \lambda^\alpha \ln (\lambda^\alpha) $ that leads to
\be
S_H \to \frac{\log_2\lambda}{\ln \lambda} -\log_2(A) = \frac{\log_2\lambda}{\ln \lambda} -\log_2\left[ \lambda^\alpha \ln (\lambda^\alpha)\right].  
\label{eq:S_turb}
\ee
} 
{Note that $\alpha = 2/3$ for hydrodynamic turbulence. For this $\alpha$, in Fig.~\ref{fig:hydro_model}(a) we plot $S_H(G)$ as a function of $G$ for  $\lambda=1.2, 1.3, 1.4$, and 1.5.  As shown in the figure, $S_H$  increases monotonically and  approaches  a constant. The asymptotic $S_H$ for $\lambda = 1.2, 1.3, 1.4, 1.5$ are  4.31, 3.71, 3.26,  2.94 (approximately) respectively.  Note that $S_H$ is nearly constant for  $G > 20$. This is because $p_n$ drops very sharply with $n$, implying that $S_H$ gets maximal contributions from small $n$'s or large-scale structures only.  Note the marginal variations in $S_H$ with $\lambda$. Hence, the choice of $\lambda$ is arbitrary, as in the shell model~\cite{Gledzer:DANS1973,Biferale:ARFM2003}.   
}

\begin{figure}[h]
	\centering
	\includegraphics[width=1\columnwidth]{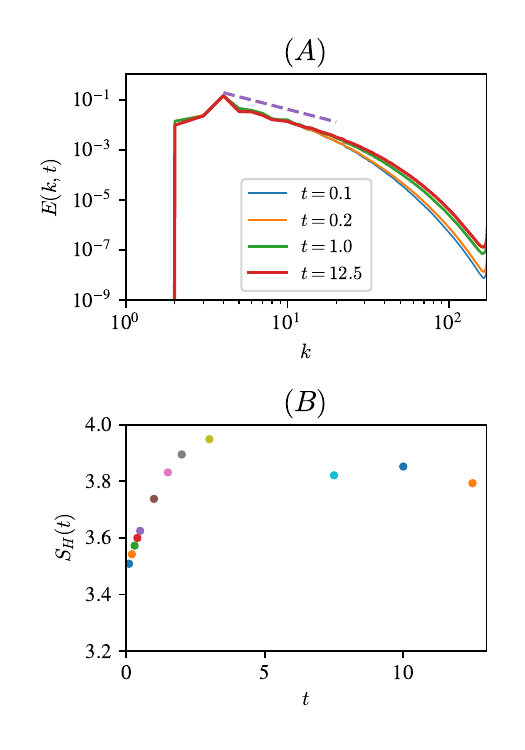} 
	\caption{For a forced hydrodynamic turbulence simulation on a $512^3$ grid: (A) Plots of $E(k)$ vs.~$k$  at  $t=0.1, 0.2, 1.0$ and 12.5 eddy turnover times using colored lines with increasing thicknesses. The dashed purple line in the figure represents $k^{-5/3}$ spectrum. The figure shows that $E(k)$ at large $k$'s grow with time. (B) Plot of the hydrodynamic entropy $S_H(t)$ vs.~$t$  exhibits increasing entropy  till $t=5$, after which $S_H$ flattens out (to $\approx 3.8$) because the flow has reached a steady state. } 
	\label{fig:hydro_num}
\end{figure}

{We compare the analytical estimates derived above with $S_H$ computed using  simulation data.   Using  pseudospectral code TARANG~\cite{Chatterjee:JPDC2018}, we simulate hydrodynamic turbulence on a periodic $(2\pi)^3$ box with $512^3$ grid points. We choose the nondimensional kinematic viscosity as $\nu = 10^{-3}$.  Using the scheme described in \citet{Sadhukhan:PRF2019}, we employ random force  at low wavenumbers with   kinetic-energy supply rate  $\epsilon=0.1$, and zero  kinetic-helicity supply rate. We time advance the Fourier modes using fourth-order Runge-Kutta (RK4) method with a time-step  $dt=10^{-3}$.  For the  time stepping, we absorb the viscous term using exponential integrating factor~\cite{Canuto:book:SpectralFluid,Verma:Pramana2013tarang}.

We start our simulation with a turbulent initial condition given by Pope~\cite{Pope:book}. After around 7  eddy turnover times, the system reaches a steady state, whose Reynolds number Re based on system size is approximately 5580.  The Reynolds number based on Taylor’s microscale  ($\mathrm{Re}_\lambda$) is  approximately 230. Our simulation is well resolved because $k_\mathrm{max} \eta \approx 1.25$, where $\eta = (\nu^3/\epsilon)^{1/4}$ is the Kolmogorov length. Figure~\ref{fig:hydro_num}(A) exhibits the energy spectrum $E(k)$ for $t=0.1, 0.2, 1.0$, and 12.5  eddy turnover times. In the figure, a narrow wavenumber range [$k=(5,30)$] exhibits $k^{-5/3}$ spectrum, which is shown as a purple dashed curve. 

As shown in Fig.~\ref{fig:hydro_num}(A), $E(k)$  at higher wavenumbers grow initially, after which $E(k)$ saturates to a steady profile~\cite{Pope:book}. The above increase in $E(k)$ at large $k$'s leads to an increase in $S_H$  in the early stages [Fig.~\ref{fig:hydro_num}(B)]. However, $S_H$ saturates to a near constant value ($\approx 3.8$) after $t=7.5$ eddy turnover times. The asymptotic $S_H$ for the simulation is near $S_H$ for $\lambda = 1.3$. A cautionary remark, the definitions of probability ($p_n$) for the analytical and numerical calculations are somewhat different. However, the sums leading to $S_H$ yield nearly the same results because all the modes have been accounted for in both  sums. The difference arising due to log-binned shells  is marginal. Note that the Boltzmann entropy computations too involve large sums where we ignore marginal terms.  }

Thus, using  analytical and numerical tools we show that the HE of hydrodynamic  turbulence is nearly constant, independent of grid size. Hence,  \textit{HE of a turbulent system is nonextensive}.  Interestingly, $S_H$ for hydrodynamic turbulence  is smaller than that for the disordered Euler turbulence [Eq.~\ref{eq:S_H_Euler}]. Hence, a turbulent flow is much more ordered than the disordered Euler turbulence. This is because the hydrodynamic structures at large scales are much more energetic than those at small scales (according to Kolmogorov's theory), which is not the case  for the disordered Euler turbulence where the energy is equipartitioned  among all Fourier modes.

For a forced turbulence, the energy spectrum $E(k)$ is maintained at a steady level. Hence, $S_H$ of a turbulent flow is approximately constant in time.  However, the  coherent energy of the fluid structures are transferred to the thermal energy at  microscopic scales, leading to an increase in the TE.  Thus, the external force applied to the fluid flow injects hydrodynamic order that is transferred to the thermodynamic entropy at small scales. However, the thermodynamic and hydrodynamic entropies have very different properties (e.g., extensivity), hence they cannot be added.    The hydrodynamic  and thermodynamic disorder of a fluid system have to be separately quantified.  We also remark that the TE undergoes a change via heat injection to the system or via work done by the system. Analogously,  hydrodynamic entropy can be altered during the decay of turbulence or via external forcing.  

In turbulence literature, a  turbulent  signal is  contrasted with a random signal using skewness, kurtosis,  and the probability distribution functions of the velocity gradients.  {We propose that the hydrodynamic entropy  can also be used  as a diagnostic tool for turbulence characterization. If the $S_H$ of a purported turbulent signal is far  from 4, then that  signal is not  turbulent.}

In the next section, we compute the HE for the Ising model and $\phi^4$ theory.

\section{Hydrodynamic entropy for several  systems of statistical mechanics }
\label{sec:HE_statmech}

First we contrast the thermodynamic and hydrodynamic entropies of the Ising model.

\subsection{Ising model}
\label{sec:Ising}
Consider $N$ Ising spins each of which takes values $+\sigma$  or $-\sigma$. { The Hamiltonian of the system is 
\bea
H = -J \sum_{\la ij \ra} \sigma_i \sigma_j - B \sum_{i} \sigma_i,
\eea
 where $J$ is the coupling constant, and $B$ is the external magnetic field.} Here, the first sum is performed over the nearest neighbours, whereas the second sum is over all the spins.  For the Ising model, the thermodynamic entropy is zero at zero temperature, whereas it takes the maximum value, $S_\mathrm{max} = k_B N \ln 2$,  at infinite temperature.  At an arbitrary temperature, we compute the entropy using  Boltzmann's formulas~\cite{Huang:book:SM}.

We can also compute the hydrodynamic entropy (HE) for the Ising model as follows. The  configuration of the Ising model can be specified using Fourier transform ($\hat{\sigma}_k$), which is,
\bea
\hat{\sigma}_k & = & \frac{1}{N} \sum_j \sigma_j \exp\left[-\frac{ \sqrt{-1} (2\pi kj)}{N} \right] = |\hat{\sigma}_k| \exp(i \delta_k), \nonumber \\
\eea
where $ |\hat{\sigma}_k|$ and $\delta_k$ are, respectively, the amplitude and phase of the mode $\hat{\sigma}_k $.  After this we define 
\be
p_k = \frac{|\hat{\sigma}_k|^2}{\sum_k|\hat{\sigma}_k|^2 },
\label{eq:p_k_hydro}
\ee
and 
\be
S_H = -\sum_k p_k \log_2 p_k.
\ee
In the ground state, all the spins are  either up $(\sigma_i =  1)$ or down $(\sigma_i =  -1)$ that leads   $ p_{k=0}=1$, and 0 for all other wavenumbers. Hence,  $S_H=0$ for the ground state.  At infinite temperature, the spins are random and uncorrelated, for which
\be
p_k = \frac{|\hat{\sigma}_k|^2}{\sum_k|\hat{\sigma}_k|^2 } = \frac{1}{N} \mathrm{~for~} k \ne 0,
\ee
and
\be
S_{H, \mathrm{max}} = \log_2 (N),
\ee
which is the maximum possible value for $S_{H}$.  At an intermediate temperature, $0< S_H<\log_2 (N)$.

Similar to Euler and hydrodynamic turbulence, $S_H$ for the Ising system is nonextensive because $S_{H, \mathrm{max}}  \propto \log(N)$, not $N$.  This is one of the major differences between TE and $S_H$. The hydrodynamic entropy captures \textit{multiscale} (at different length scales) order, whereas thermodynamic entropy captures \textit{microscopic order}.  Interestingly, the energy spectrum $E({\bf k})$ is independent of the phase of the mode ${\bf u}({\bf k})$. Thus, the HE is constructed using less information of the system, yet it captures the random nature of the steady state quite well.  This is one reason why $S_H$ is less than the corresponding BE.

We remark that for the Ising model, the Shannon entropy and $S_H$  are different even though their formulas are similar. As an example, for the maximally random Ising model,  $p_{1} = p_{-1} = 1/2$ for every spin that leads to 
\be
S_{\mathrm{Shannon}}  = N(-p_1 \log_2 p_1 -p_{-1} \log_2 p_{-1}) =  N,
\ee
which is very different from $S_H$, which is $\log_2 (N)$.   

Next, we compute the hydrodynamic entropy for the $\phi^4$ theory.

\subsection{$\phi^4$ theory and related equations}
	
The \textit{time-dependent Ginzburg-Landau} (\textit{TDGL})  that describes evolution of order parameter $\phi$ of $\phi^4$ theory is~\cite{Chaikin:book}
\be
\partial_t \phi = \phi - \phi^3 + \nabla^2 \phi + \eta,
\label{eq:TDGL}
\ee
where $\eta$ is the noise that simulates  heat bath. The above system, called \textit{Model A}, conserves $\int dx \phi(x)$. This equation is used for describing many physical phenomena, e.g., phase separation, phase transition, superconductivity, etc. It is important to note that TDGL equation is, in some sense, a hydrodynamic description of microscopic systems, such as spins.

An  equilibrium solution of the TDGL equation is random $\phi$ with a zero mean. This solution is analogous to the random Ising spins described in Sec.~\ref{sec:Ising}, and its HE is $\log_2 N$, where $N$ is the number  of grid points used for discretizing Eq.~(\ref{eq:TDGL}). TDGL equation also admits two ordered equilibrium solutions, $\phi = 1$ and $-1$, whose hydrodynamic entropies are zero.

TDGL equation has two nonequilibrium steady-state solutions---kink and antikink---that approach +1 at one end and $-1$ at the other end. These solutions are  $\pm \tanh[(x-x_0)/\sqrt{2}]$, with size of the domain walls of the order of unity~\cite{Chaikin:book}.  The evolution of $\phi$ to the kink or anti-kink or the asymptotic state, $\phi(x)=\pm 1$, are dynamic and out of equilibrium, for which we may define  hydrodynamic entropy. 

\begin{figure}[h]
	\centering
	\includegraphics[width=1\columnwidth]{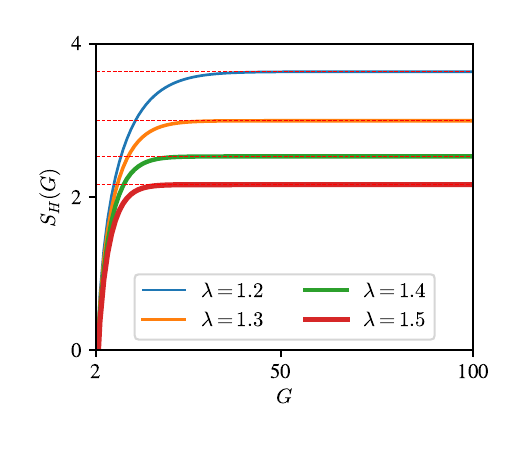} 
	\caption{For kink-antikink pairs of coarsening, plots of hydrodynamic entropy $S_H(G)$ vs.~$G$ computed using  the analytical formula [Eq.~(\ref{eq:S_H_anlytical})] with  $\alpha = 1$ and $\lambda=1.2, 1.3, 1.4, 1.5$  using colored lines of increasing thicknesses. The asymptotic values [Eq.~(\ref{eq:S_turb})] for the four  $\lambda$'s are 3.64, 2.99, 2.53, and 2.16 respectively (horizontal dashed lines).  }
	\label{fig:coarsen_model}
\end{figure}

Following Porod's law~\cite{Porod:book_chapter,Verma:PRE2023_coarsening}, the power spectrum of the kink and antikink are
\be
E(k) = \frac{\mu^2}{L} k^{-2},
\ee
where $\mu$ is the jump in $\phi$ across the two phases, and $L$ is the system size. {The above spectrum  arises due to the discontinuity in $\phi(x)$ at the interface, and it is valid in dimensions 1, 2, and 3. 	 Hence, the SH formula to be derived below is applicable in any dimension, as long as the $\phi(x)$ contains a kink or anti-kink. }
Following the same arguments as in Sec.~\ref{sec:HE_turbulent}, we can show that for such a  system,
\be
p_n = \frac{E_n}{E} = A (k_n L)^{-1} = A \lambda^{- n},
\ee
and the hydrodynamic entropy is given by Eq.~(\ref{eq:S_H_anlytical}) with $\alpha =1$. See Fig.~\ref{fig:coarsen_model} for an illustration of $S_H(G)$ for  $\lambda = 1.2, 1.3, 1.4, $ and 1.5.  For large $G$, $S_H$ of a coarsening system is 
\be
S_H \approx  \frac{\log_2\lambda}{\ln \lambda} -\log_2(A) = \frac{\log_2\lambda}{\ln \lambda} -\log_2\left[ \lambda \ln (\lambda)\right].  
\label{eq:S_coarsening}
\ee
{The asymptotic values of $S_H$ for $\lambda=1.2, 1.3, 1.4, 1.5$  are 3.64, 2.99, 2.53, and 2.16  respectively, which are presented as dashed red lines in Fig.~\ref{fig:coarsen_model}. In addition, the TDGL equation admits a random state, whose $S_H = \log_2 N$, and two other asymptotic states, $\phi = \pm 1$, for whom   $S_H = 0$. Note that the maximum TE for the random state is $O(N)$. Hence, the HE of the TDGL equation is nonextensive because it is never proportional to $N$.   }

\begin{figure}[h]
	\centering
	\includegraphics[width=1\columnwidth]{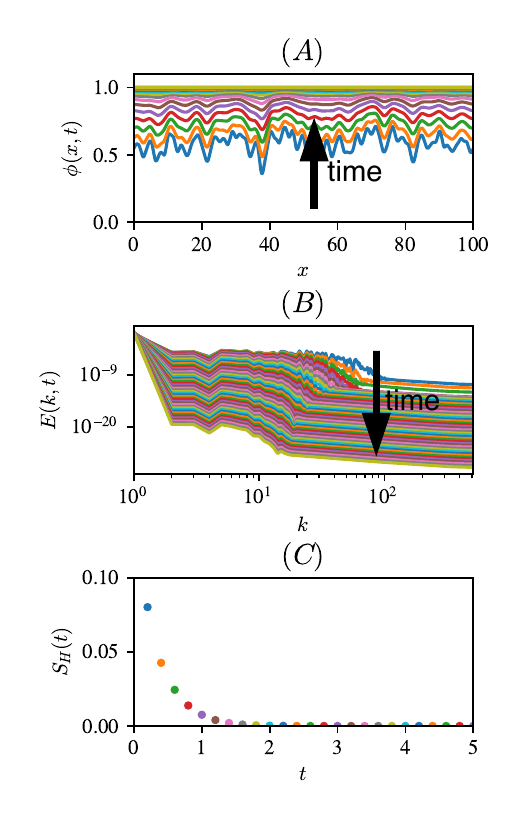} 
	\caption{For a simulation of TDGL equation with biased initial condition ($\la \phi(x,t=0) \ra \ne 0$): (A) Plots of $\phi(x,t)$ vs.~$x$ from $t=0.1$ to 9.8. $\phi(x,t)$ starts from the blue-colored configuration and increases to $\phi(x,t) = 1$ near $t \approx 3$.  (B) The respective energy spectrum $E(k,t)$ vs.~$k$; $E(k,t)$  spreads out initially, but it tends to concentrate at lower $k$'s.  Asymptotically, $E(k) $ is nonzero only for $k=0$. (C) Plot of hydrodynamic entropy $S_H(t)$  up to $t=5$ shows that $S_H(t)$  decreases asymptotically to zero.  }
	\label{fig:coarsen_num}
\end{figure}

{We illustrate the evolution of HE using a direct numerical simulation of TDGL equation. We analyze the numerical data generated by \citet{Verma:PRE2023_coarsening}, who simulated the TDGL equation using finite difference method in a one-dimensional (1D)  periodic box of size $L=100$ with a grid resolution of 1024. They took the time step $\Delta t = 0.001$ that satisfies the Courant–Friedrichs–Lewy (CFL) or linear-stability condition~\cite{Verma:PRE2023_coarsening}. They evolved a biased initial state ($\la \phi(x, t=0) \ne 0 \ra$) from $t=0$ to $t=9.8$. In Fig.~\ref{fig:coarsen_num}(A)  we illustrate the evolution of $\phi(x, t)$ from a random state (blue curve) to the asymptotic state, $\phi(x) = 1$. 

For the above $\phi(x,t)$, the corresponding energy spectra ($E(k)$) and hydrodynamic entropies ($S_H(t)$) are exhibited in Fig.~\ref{fig:coarsen_num}(B,C) respectively. As shown in Fig.~\ref{fig:coarsen_num}(B), the initial $E(k,t)$ are spread out among a wide range of wavenumbers. With time, $E(k)$ for large $k$'s dissipates more strongly than those for small $k$'s~\cite{Verma:PRE2023_coarsening}. Consequently, $E(k,t)$ becomes more skewed towards small wavenumbers. Hence, the HE of the system decreases monotonically with time, as illustrated in Fig.~\ref{fig:coarsen_num}(C). The HE of the asymptotic state is zero. 

For an unbiased initial condition, the TDGL equation evolves to one or more kink-antikink pairs. The HE for the asymptotic state would be given by Eq.~(\ref{eq:S_coarsening}).  Following similar analysis as above, we can numerically verify these results.  In addition, we can also compute the hydrodynamic entropy for model B~\cite{Puri:book_edited}, and expect the behaviour to be similar.  To maintain focus on the main theme, we do not present these results here. 
}

We also remark that the emergence of ordered states---$\phi = \pm 1$ or kink/antikink---in 1D TDGL equation does not contradict  \textit{Mermin-Wagner theorem}, according to which long-range order (LRO) does not exist in 1D and 2D \textit{equilibrium} systems with short-range interactions.  Nonequilibrium nature of  turbulent flows and TDGL equation invalidates the arguments of Mermin-Wagner theorem.  Note that Toner and Tu~\cite{Toner:PRE1998} observed flocking and related long-range ordered states in 2D active matter.


The above  examples contrast the hydrodynamic and thermodynamic entropies. In the next section, we summarize our results.

\section{Discussions and Conclusions}
\label{sec:conclusions}

We start this paper with   \citet{Prigogine:Science1978}'s
motivation for the  emergence of the dissipative structures in Rayleigh-B\'{e}nard convection.  In this paper we argue that such convective patterns, which are more ordered than conduction profile, require new measures for disorder quantification (other than thermodynamic entropy). 
In this paper, we show that hydrodynamic entropy (HE) captures such a multiscale order/disorder. 

In this paper, we compute the hydrodynamic entropies of Euler and hydrodynamic turbulence, TDGL equation, and Ising spins.  Via these examples, we show that the HE is nonextensive, and it differs significantly from the thermodynamic entropy (TE). We summarize  some of the important properties of  HE as follows:

\begin{enumerate}
	\item  HE is a measure of multiscale and macroscopic disorder. In contrast, TE captures microscopic disorder (single-scale phenomena). However, there is an exception: TE capture multiscale disorder near  second-order phase transition.

	\item TE is an extensive quantity, that is, it depends on the system size. However,  HE is nonextensive. For example, the HE of equilibrium Euler turbulence  is $\log_2 M$, where $M$ is the grid size. The HE of a turbulent flow is even lower (between 3 and 4), and it is dominated by the large-scale structures.

	\item Since TE and HE have very different properties, we cannot add them to compute the total entropy of a system.  We use the respective entropies to specify the disorder at different scales. This is in the same spirit why we ignore quantum fields and fundamental particles---quarks and electrons---while describing gases and liquids.  
\end{enumerate}

Many natural processes are out of equilibrium and multiscale.  Some of the popular examples are plasma turbulence~\cite{Matthaeus:APJ2020}, financial systems~\cite{Zhou:Entropy2013}, ecological systems, electroencephalogram (EEG)~\cite{Deco:Cell_Reports2020}, electrocardiogram (ECG)~\cite{Asgharzadeh:Bio2020}, voltage-dependent anion chanels~\cite{Verma:EPL2006}, etc. {These systems exhibit power-law spectrum, somewhat similar to turbulence. Hence, we expect their HE to have similar properties. Tsallis entropy~\cite{Tsallis:book} appears to describe order of   these systems satisfactorily (see Appendix A).  For quantum systems, quantification of order  and entanglement  remains a challenge, and it is an active area of research. von Neumann entropy, entanglement entropy, and information entropy are popular  measures in this field~\cite{Zurek:RMP2003} (see Appendix A). For quantum systems, construction of hydrodynamic entropy, and its comparison with other entropies will be an interesting exercise. 
	
Among the entropies discussed in Appendix A, only Tsallis entropy is nonextensive, and the rest are extensive.  Note that HE too is nonextensive. Hence, a detailed comparison between the Tsallis entropy and HE will be useful. In addition, the formula for HE, Eq.~(\ref{eq:S_H}), resembles Shannon entropy formula.  But, HE and Shannon entropy are very different, as we illustrate in Sec.~\ref{sec:Ising}.  This paper brings out similarities and dissimilarities between various entropies. Still, further   comparison between  them is required.}

{ Researchers often employ particle-based simulations, e.g, molecular dynamics (MD),  to study microfluids, chemical reactions, turbulence, and interacting many-particle systems~\cite{Bird:book}. Thermodynamic entropy is useful for such systems when they are in equilibrium or near equilibrium. However, TE does not capture  the order accurately in  MD simulations involving thermal convection, turbulence~\cite{Gallis:PRL2017},  granular matter~\cite{deGennes:RMP1999},  or strongly-interacting plasmas~\cite{Wani:PP2024}. We believe that HE computed using coarse-grained hydrodynamics may be useful  for quantifying order in such systems.  We hope that such exercise will be performed in future.}

Nonequlibrium systems typically exhibit energy transfers across scales. In hydrodynamic turbulence, the multiscale energy transfer is proportional to the third-order structure function or triple-order correlation of Fourier modes~\cite{Kolmogorov:DANS1941Structure,Pope:book,Verma:book:ET}. Hydrodynamic entropy ignores the phases of the Fourier modes, hence it can not provide  information on the energy transfers or phase correlations.    The  \textit{phase entropy}~\cite{Kuramoto:book}, which is used for quantifying \textit{phase synchronization},  may be useful for the analysis and global measure of  multiscale energy transfers.

We  emphasize that the multiscale framework provides valuable insights into the noequilibrium systems. For example, multiscale energy transfers can be employed to break the time reversal symmetry in nonequilibrium driven systems~\cite{Verma:EPJB2019,Verma:PTRSA2020}.  Hydrodynamic entropy,  an important quantity in the multiscale framework,  could complement past works in this field.


\vspace{0.3cm} 
{\em Acknowledgements }: The authors thanks Riddhi Bandyopadhyay, Anurag Gupta, and Adhip Agarwala for useful discussions; and Abhishek Jha and Pradeep Yadav for providing  simulation data. MKV gratefully acknowledges the support from Science and Engineering Research Board, India,  for the J. C. Bose Fellowship (SERB /PHY/2023488) and for the grants SERB/PHY/20215225 and SERB/PHY/2021473.  RS thanks IIT Kanpur for the visiting faculty position in 2023 and 2024, during which this work was done.

\appendix
\section{Various Entropies Used in Physics}
\label{sec:Appendix}

In this Appendix, we summarize  thermodynamic entropy, Shannon entropy, von Neumann entropy, and Tsallis entropy. In the main text, we compare these entropies with the hydrodynamic entropy.

\vspace{0.3cm}

{\em Thermodynamic entropy}:  In statistical mechanics, if a system has $W$ microstates or configurations that occur with equal probability, then  the \textit{configuration entropy} or \textit{Boltzmann entropy} is given by $S  = k_B \log W$, where $k_B$ is Boltzmann constant~\cite{Landau:book:StatMech}.   At zero temperature ($T$),   $W=1$ and $S=0$, which is the minimum entropy for the system.  However, at infinite temperature, the system is completely random and the entropy is maximal.  For the random Ising system with $N$ spins,  $W=2^N$ and $S = k_B \log W= k_B N \log 2$.  

The thermodynamic entropy of an ideal gas with $N$ particles in a box of volume $V$ with a total energy of $E$  is~\cite{Huang:book:SM}
\be
S = N k_B \left[ \log \frac{V}{N} + \frac{3}{2} \log \frac{4\pi m E}{3Nh^2}  + \frac{5}{2} \right],
\label{eq:S_thermo_gas}
\ee
where $h, k_B$ are the Planck and Boltzmann constants, and $m$ is the mass of each particle. The above thermodynamic entropies are extensive because they are proportional to $N$.

According to Gibbs~\cite{Gibbs:book:StatMech}, a macrostate of a thermodynamic system is combination of microstates. If probability of $i^\mathrm{th}$ microstate is $p_i$, then the Gibbs  entropy is 
\be
S_G = -k_B \sum_i p_i \log p_i.
\ee
At a finite temperature, $S = -\partial F/\partial T$ where $F = -k_B T \log Z$ with  $Z = \sum_C \exp(-\beta  E_C)$.  Since $\log Z $ is  proportional to the system size, the  thermodynamic entropy $S$ is extensive~\cite{Huang:book:SM}.

\vspace{0.3 cm}

{\em Shannon entropy}: Shannon entropy~\cite{Shannon:BELL1948} is very useful in computer science and electrical engineering, in particular for data compression and transmission.   Imaging an string of $n$ random symbols, where the random symbol $x$ occurs with probability  $p_x $. Then, the Shannon entropy  is defined as
\be
S = - \sum_x  p_x \log_2 p_x .
\ee
 If $x$ has $n$ possible outcomes that are equally probable, then $p_x = 1/n$ for all $n$, and 
\be
S = S_\mathrm{max} =- \sum_x  \frac{1}{n} \log_2  \frac{1}{n}  = \log_2(n) .
\ee 
When the probabilities for the outcomes are unequal, it is easy to show that $S < S_\mathrm{max}$.   Note that the Shannon entropy is extensive. 

\vspace{0.3 cm}

{\em von Neumann entropy}: von Neumann entropy~\cite{Zurek:RMP2003}, an extension of Gibbs entropy, is useful for entropy computation  of quantum systems. It is
\be
S = -\mathrm{Tr}(\hat{\rho} \log \hat{\rho}),
\ee
where $\hat{\rho} $ is the density matrix,
\be
\hat{\rho} = \eta_j |j><j|
\ee
with $|j>$ as the basis state.  When $|j>$'s are the eigenvectors of the system, then 
\be
S = -\sum_j \eta_j \log(\eta_j),
\ee 
which is same as the Gibbs entropy. 

For a quantum system with $N$ qubits  of spin 1/2, the Hilbert space is $2^N$ dimensional. If each quantum state has equal probability, then $\eta_j = 2^{-N}$ for all $j$'s that leads to
\be
S = N \log 2,
\ee
which is the maximum value of the von Neumann entropy. Since $S \propto N$, the von Neumann entropy is extensive. 

\vspace{0.3 cm}

{\em Tsallis entropy}: Tsallis entropy~\cite{Tsallis:book}, $S_q$, an extension of Boltzmann and Gibbs entropies, provides expectation of q-logarithm of a distribution:
\be
S_q =k  \frac{1-\sum_{i=1} p_i^q }{q-1}   
\ee
where $\sum_i p_i = 1$, $q$ is a real parameter, and $k$ is a positive constant. When $q \rightarrow 1$, $S_q \rightarrow -k \sum_i  p_i \log_2 p_i $, which is Gibbs entropy.  Note that Tsallis entropy is not additive, and hence nonextensive (except when $q \to 1$). Tsallis entropy finds application in chaos theory, dusty plasma, spin glass relaxation, etc. { Note that   Tsallis entropy measures disorder at single-scale, whereas hydrodynamic entropy captures multiscale disorder. Still, it is important to compare the nonextensive features of the two entropies.

The other important entropies employed in physics are entanglement and information entropies~\cite{Zurek:RMP2003}, black hole entropy~\cite{Giddings:PT2013}, and phase entropy~\cite{Kuramoto:book}. However, we do not discuss them here because they are beyond the scope of this paper.}
  

%

 
\end{document}